\documentclass[letter]{aa}
\usepackage{natbib,twoopt}
\usepackage[breaklinks=true]{hyperref}
\usepackage{graphicx}
\usepackage{txfonts}
\usepackage{lipsum}
\usepackage{subcaption}
\usepackage{lscape}
\usepackage{placeins}

\usepackage{makecell, booktabs, amsmath, tabularx}

\newcommand{\ergs}{\rm erg\ s^{-1}}
\newcommand{\lr}{L_{\rm R}}
\newcommand{\lx}{L_{\rm X}}
\newcommand{\ledd}{L_{\rm Edd}}

\newcommand{\msun}{M_{\odot}}

\begin{document}

   \title{Detection of a parsec-scale, compact, and fading ejecta from an accreting massive black hole}

   \author{Chao Li\inst{1,2}\email{lichao@xao.ac.cn}
        \and Ning Chang\inst{1}\email{changning@xao.ac.cn}
        \and Jun Yang\inst{3}\corrauth{jun.yang@chalmers.se}
        \and Lang Cui\inst{1,2,4}\corrauth{cuilang@xao.ac.cn}
        \and Luis C. Ho\inst{5,6}\email{lho.pku@gmail.com}
        }

   \institute{State Key Laboratory of Radio Astronomy and Technology, Xinjiang Astronomical Observatory, CAS, 150 Science 1-Street, Urumqi, Xinjiang, 830011, P. R. China
   \and School of Astronomy and Space Science, University of Chinese Academy of Sciences, No.1 Yanqihu East Road, Beijing 101408, P.R.China
   \and Department of Physics and Astronomy, Chalmers University of Technology, Onsala Space Observatory, SE-43992 Onsala, Sweden
   \and Xinjiang Key Laboratory of Radio Astrophysics, 150 Science 1-Street, Urumqi 830011, P.R.China
   \and Kavli Institute for Astronomy and Astrophysics, Peking University, Beijing 100871, P.R.China
   \and Department of Astronomy, School of Physics, Peking University, Beijing 100871, P.R.China}

   \date{Received April 15, 20XX}

\abstract 
{Dwarf galaxies, characterized by their low luminosities and masses, are excellent candidates for searches for intermediate-mass black holes (IMBHs), particularly when they show strong accretion and ejection activity. The dwarf galaxy SDSS J101747.09+393207.7 has recently been found to display a very high X-ray luminosity and an X-shaped optical structure, possibly caused by a dwarf--dwarf merger. To explore its potential IMBH ejection activity, we performed very long baseline interferometry (VLBI) observations at 4.9 GHz. In this work, we present the detection of a milliarcsecond-scale, compact, sub-microjansky radio component near the optical centroid. According to some existing radio sky survey data, the radio component was not detected until 2015; it displayed an optically thin steep radio spectrum and declining flux densities across 0.8--5 GHz from 2019 to 2025. Therefore, we identify it as a short-lived and rarely seen ejecta that was produced by unstable accretion onto a massive black hole and likely faded away in a few decades. These results indicate that short-lived, episodic jet activity from accreting IMBHs in dwarf galaxies might exist.}


   \maketitle
   \nolinenumbers
   
\section{Introduction}\label{sec:1}

Supermassive black holes (SMBHs) have been widely observed in the very early Universe \citep[e.g.,][]{2024Natur.627...59M,2024Natur.636..594J,2025arXiv251203130I}, but their formation and growth mechanisms remain unclear. Galaxy mergers represent one potential trigger for active galactic nuclei (AGNs) and the subsequent rapid accretion of black holes \citep[e.g., the possible $z\sim11$ merger reported by][]{2023ApJ...949L..34H}. Direct observations of high-redshift mergers are challenging, making local dwarf galaxies critical analogs. Investigating dwarf--dwarf interactions and black hole accretion may thus elucidate the early evolution of SMBHs \citep[e.g.,][]{2024ApJ...968L..21M}.

Dwarf galaxies, typically defined by stellar masses of $M_\star = 10^{6.0-9.5}~M_\odot$ \citep[see reviews by][]{2020ARA&A..58..257G}, represent the most common galaxy population in the Universe. A subset of these systems host elusive intermediate-mass black holes \citep[IMBHs; $M_{\rm BH} = 10^{2-6}~M_\odot$;][]{2020ARA&A..58..257G}, which are thought to be the progenitors of SMBHs. Recently, \cite{2025ApJ...982...10P} identified a substantial sample of 2444 AGN candidates (2.1\%) in dwarf galaxies using early data from the Dark Energy Spectroscopic Instrument (DESI) survey. However, the intrinsic faintness of these AGNs \citep[particularly at X-ray wavelengths;][]{2024A&A...681A..97A} introduces significant selection biases across optical, infrared, and X-ray detection methods \citep[e.g.,][]{2024ApJ...971...68W,2025arXiv250109791S}, complicating demographic studies. In the radio band, jets are one of the most direct tracers of IMBH accretion activity. Although arcsecond-scale observations successfully identify AGNs in dwarf galaxies \citep[e.g.,][]{2020ApJ...888...36R,2025ApJ...978..158E}, higher resolution is required to isolate the compact nonthermal emission. To date, only six IMBH candidates with radio counterparts have been detected with very long baseline interferometry (VLBI); all six show parsec-scale nonthermal structures from episodic jets \citep[e.g.,][]{2023MNRAS.520.5964Y}. None of these dwarf galaxies shows signs of mergers or strong interactions. 

Cosmological simulations show that dwarf--dwarf mergers dominated the galaxy merger rate in the early Universe. However, owing to instrumental sensitivity limitations, observational evidence of merger-triggered AGN activity has thus far been restricted to nearby systems. Recently, \cite{2024ApJ...968L..21M} demonstrated that grouped or paired dwarf galaxies exhibit X-ray AGN detection rates elevated by a factor of 6-10 compared to isolated counterparts, providing unprecedented empirical constraints on cosmological models. 

\begin{figure*}[htbp]
    \centering
    \includegraphics[height=0.38\linewidth]{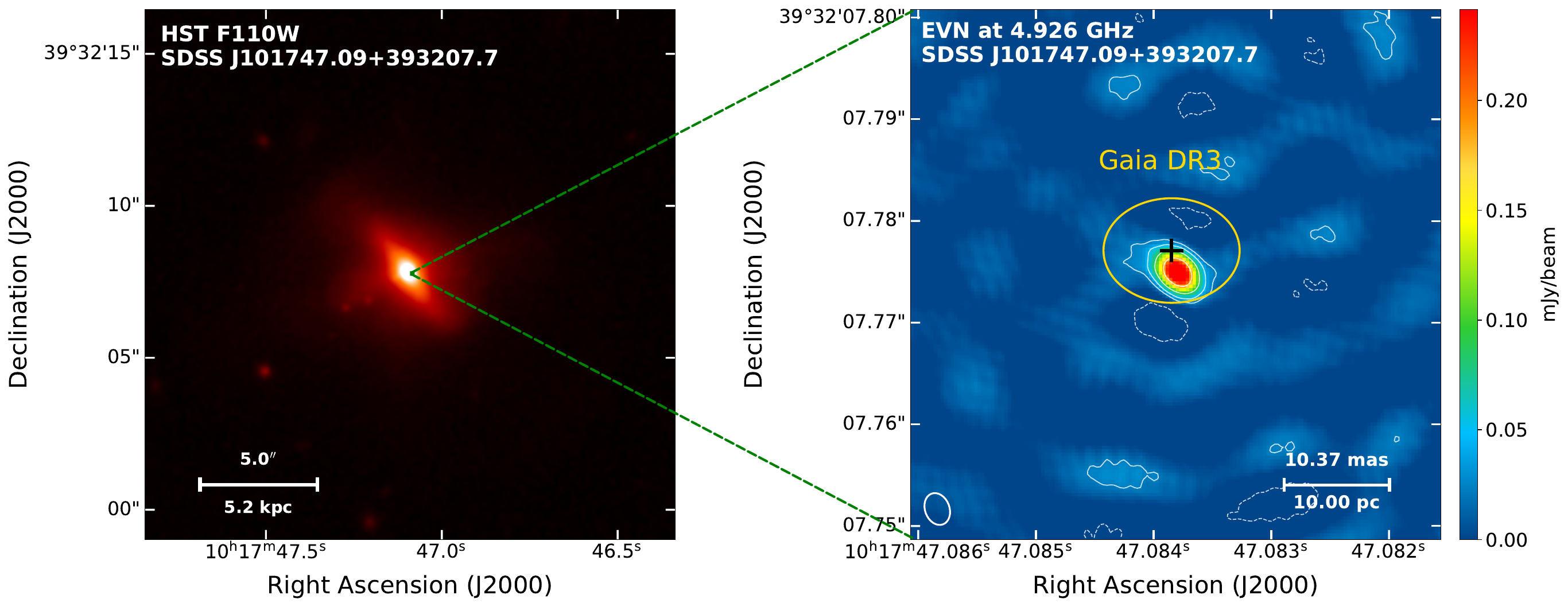}
    \caption{Near-infrared and EVN observed images of J1017. \textit{Left}: HST observation of J1017 with the infrared filter F110W. \textit{Right}: 4.926 GHz VLBI image of J1017 with a beam size (FWHM) of $\rm 3.2\times2.4~mas$ and a major axis position angle of $\rm 20.9^\circ$. Contours are set at $3\sigma\times(-1,1,2,4,8)$. A scale of $10.37\ \mathrm{mas}$ corresponds to a physical distance of $10\ \mathrm{pc}$. The black cross marks the optical position from \textit{Gaia} DR3; the semimajor and semiminor axes of the yellow ellipse represent the total $\rm 1\sigma$ uncertainties. The dashed green line indicates that the VLBI imaging field of view corresponds to the marked region in the near-infrared image.}
    \label{fig:1}
\end{figure*}

SDSS 101747.09+393207.7 (hereafter J1017) is a local dwarf galaxy with a redshift of $z=0.054$. Its star formation rate is $0.34\ \msun$ yr$^{-1}$ \citep{2024ApJ...974...51K}. The left panel of Fig. \ref{fig:1} displays a high-resolution near-infrared image from the \textit{Hubble} Space Telescope (HST) using the F110W filter, which reveals an X-shaped morphology indicative of tidal features from gravitational interaction; such structures appear without an obvious companion galaxy in the vicinity, suggesting a late-stage merger near the point of coalescence \citep{2021ApJ...911..134K,2024ApJ...974...51K}. This system displays an unusually high X-ray luminosity of $L_{\rm 2-10~keV} = 10^{42.2}~\rm erg~s^{-1}$, an unambiguous signal of AGN activity \citep{2024A&A...681A..97A,2024ApJ...974...51K}. Using the $M_{\rm BH}$--$\sigma_\star$ relation, its black hole mass is estimated to be $10^{5.3\pm0.55}~M_\odot$ \citep{2024ApJ...974...51K}. With a stellar mass of $\sim$$10^{9.0}~M_\odot$, J1017 hosts a rare compact optical nucleus with a half-light radius of merely $0.4$~kpc \citep{2024ApJ...974...51K}, making it a critical laboratory for studying black hole growth in low-mass galaxy mergers.

This paper is organized as follows. We describe the European VLBI Network (EVN) observations and data reduction in Sect. \ref{sec:2}. Section \ref{sec:3} presents our deep EVN imaging results. We discuss component properties and implications in Sect. \ref{sec:4}, and give our conclusions in Sect. \ref{sec:5}. A standard $\Lambda$ cold dark matter cosmological model with $H_0 = 71\ \mathrm{km}\ \mathrm{s}^{-1}\ \mathrm{Mpc}^{-1}$, $\Omega_{\rm m} = 0.27$, and $\Omega_{\Lambda} = 0.73$ is adopted throughout this work.

\section{VLBI observations and data reduction}\label{sec:2}

The EVN observation of J1017 at 4.926 GHz was conducted for $\sim12$ h from 2025 May 13 to 14. The participating telescopes were Jodrell Bank Mk2 (Jb2), Westerbork (Wb), Effelsberg (Ef), Onsala (O8), Tianma (T6), Toru$\rm \acute{n}$ (Tr), Yebes (Ys), Hartebeesthoek (Hh), Irbene (Ir), and Sardinia (Sr). All EVN telescopes adopted a standard observing setup (4 subbands with 4 polarizations and 64 spectral channels each, 32 MHz bandwidth per subband, 2-bit quantization, 1 second integration time). The observation was performed in phase-referencing mode with the phase calibrator J1014+3953 (separation of 46 arcmin, correlation amplitude of $\sim$70 mJy at 4.926 GHz, a nearly point-source structure). The correlation was performed by the EVN software correlator \citep[SFXC;][]{2015ExA....39..259K} at the Joint Institute for VLBI, ERIC (JIVE) using standard correlation parameters of continuum experiments.

The visibility data were calibrated using the National Radio Astronomy Observatory (NRAO) Astronomical Image Processing System (\textsc{aips}) software package \citep[version 31DEC25;][]{2003ASSL..285..109G}. Instrumental delay and bandpass calibration were performed using the bandpass calibrator J1023+3948. Global fringe-fitting was carried out using the phase calibrator J1014+3953, together with the two bandpass calibrators J1023+3948 and J0555+3948. For T6, which suffered from a low recording rate and network issues, a two-round \textsc{difmap} imaging of J1023+3948 was performed: first with T6 flagged to obtain a clean model, then with T6 included and self-calibration applied to produce the final model, which was used for subsequent amplitude calibration in \textsc{aips}. For Wb, channels 20--35 of the seventh intermediate frequency were additionally flagged because of severe polarization leakage. The first 0.1 min of each Wb scan was discarded (\texttt{quack}) to mitigate system-calibration lag effects. Tr exhibited phase oscillations on timescales of $\sim$ 6 min, recurring every $\sim$ 30 min; scans with the most severe instability were fully flagged, which necessitated the calibration described above.

Phase and amplitude solutions for the phase calibrator were derived in \textsc{aips} using \texttt{calib} (without normalization), \texttt{sncor} (phase reset), and \texttt{clcal} (transferring solutions to the target). A clean model of the calibrator obtained from \textsc{difmap} was used in this process. In \textsc{difmap}, the target data were shifted by $(-15.837,\ -9.898)$ mas (east, north) to restore the brightness peak to the phase center. Final imaging of the phase calibrator and target was performed in \textsc{difmap}; no self-calibration was applied to the target.

\section{EVN imaging results}\label{sec:3}

The 4.926 GHz EVN imaging results for J1017 are shown in the right panel of Fig. \ref{fig:1}. The optical centroid, reported in \textit{Gaia} Data Release 3 \citep[DR3;][]{2016A&A...595A...1G,2023A&A...674A...1G}, is marked as a black cross (J2000, RA $=10^\mathrm{h}17^\mathrm{m}47.083847^\mathrm{s}$, Dec. $=39^\circ32^\prime07.7771^{\prime\prime}$, $\rm \sigma_{ra}=0.74\ mas$, $\rm \sigma_{dec}=0.59\ mas$). The radio centroid ({$10^\mathrm{h}17^\mathrm{m}47.083769^\mathrm{s}$, $39^\circ32^\prime07.7749^{\prime\prime}$}) is offset by 2.38 mas from the optical centroid. \textit{Gaia} DR3 lists an \texttt{astrometric\_excess\_noise} of 5.11 mas, which indicates a deviation from a point-source model \citep{Lindegren2012A&A...538A..78L} and leads to a total uncertainty of $5.16\ \mathrm{mas}$ in right ascension and $5.14\ \mathrm{mas}$ in declination. This relatively large \texttt{astrometric\_excess\_noise} reflects the extended and asymmetric optical morphology of the host galaxy and hints at a merger nature for J1017 \citep[e.g.,][]{Hwang2020ApJ...888...73H}. Furthermore, we compared the radio centroid with the X-ray centroid \citep[$10^\mathrm{h}17^\mathrm{m}47.09^\mathrm{s}$, $39^\circ32^\prime08.11^{\prime\prime}$;][]{2024ApJ...974...51K} and found a positional offset of 0.36 arcsec; this is within the absolute astrometric uncertainty of \textit{Chandra} ($\sim$ 1 arcsec), confirming their spatial coincidence.

Nonthermal emission from J1017 is detected on the milliarcsecond scale of VLBI observations. After performing the \textsc{clean} procedure in \textsc{difmap}, we adopted a circular Gaussian model for the fitting. For J1017, we derived a total flux density of $0.41 \pm \rm 0.04\ mJy$ and a peak brightness of $0.30\ \mathrm{mJy\ beam^{-1}}$, with an RMS noise level of $7.6\ \mathrm{\mu Jy\ beam^{-1}}$. The full width at half maximum (FWHM) of the circular Gaussian model is $1.92\pm0.21$ mas. The resulting image shows that the radio emission is dominated by a compact component. We estimated the average brightness temperature using \citep{1982ApJ...252..102C}
\begin{equation}
    T_{\rm B}=1.22 \times 10^9\frac{S_{\rm obs}}{\nu_{\rm obs}^2\theta_{\rm size}^2}(1+z),
\end{equation}
where $S_{\rm obs}$ is the observed total flux density (in mJy), $\nu_{\rm obs}$ is the observing frequency (in GHz), $\theta_{\rm size}$ is the FWHM of the circular Gaussian model (in mas), and $z$ is the redshift. The average brightness temperature of J1017 is $(\rm 5.9 \pm 0.6) \times 10^6\ K$. We calculated its radio luminosity at 4.926 GHz to be $(1.39 \pm 0.14) \times 10^{38}\ \mathrm{erg\ s^{-1}}$. The errors in the total flux density and $\lr$ include the systematic errors (10 percent) due to the limited amplitude calibration accuracy.

\section{Discussion}\label{sec:4}

\subsection{Nature of the compact radio emission}

The VLBI position of J1017 is consistent with both the \textit{Gaia} DR3 optical centroid (right panel of Fig. \ref{fig:1}) and the \textit{Chandra} X-ray centroid within the uncertainty. This spatial coincidence implies the radio emission traces either the jet base or a jet component associated with the central accretion system (i.e., the accretion disk and corona) of the central IMBH. The brightness temperature of $>10^{6}$ K further confirms its nonthermal nature.

Regarding the possibility of a supernova or supernova remnant, although the observed radio luminosity falls within the range observed for luminous young radio supernovae \citep[e.g.,][]{2002ARA&A..40..387W}, this scenario is unlikely due to multiple constraints. First, following \cite{2011ApJ...737...67M}, the star formation rate of $0.34\ \msun$ yr$^{-1}$ leads to an expected core-collapse supernova rate of $\sim0.004$ yr$^{-1}$. Archival survey limits constrain any such recent transient event to a maximum time window of 30 years (see Table \ref{tab:survey-flux}). The Poisson probability of a supernova occurring within this time window is $P(N\geq1)=1-e^{-(0.004\times30)}\approx0.11$, corresponding to a low significance of $\sim 1\sigma$. The unusually high 2-10 keV X-ray luminosity of $\lx=10^{42.2}\ \ergs$ \citep{2024ApJ...974...51K} more than $\sim8$ years after the first radio detection (see Table \ref{tab:survey-flux}) also disfavors this scenario. Therefore, the compact radio emission is most naturally explained as an accretion-driven event associated with the central IMBH. Given the post-merger environment of J1017, this radio activity could be driven by accretion rate variations resulting from merger-driven gas inflows, though a tidal disruption event cannot be ruled out.

\begin{figure}[]
    \centering
    \includegraphics[width=1\linewidth]{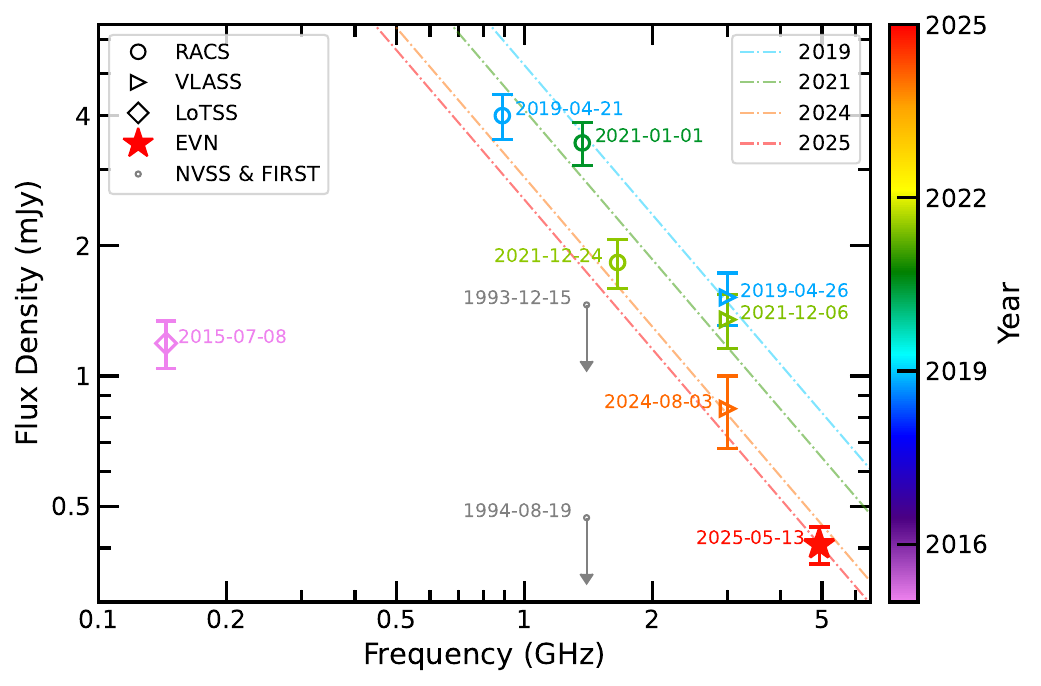}
    \caption{Radio spectra of J1017 from nonsimultaneous observations. The dashed lines show the temporal spectrum model $S(\nu,t)$. The color of these lines and the data points denotes the year. Gray points represent $3\sigma$ upper limits of NVSS and FIRST.}
    \label{fig:2}
\end{figure}

Young jets are extremely compact \citep[typically on parsec scales or even smaller; e.g.,][]{2021A&ARv..29....3O} and are expected to show spatial coincidence with both optical and X-ray centroids. Based on this together with the arcsecond-scale steep spectrum (Fig. \ref{fig:2}) and brightness temperature, we interpret the emission as an optically thin young jet rather than a self-absorbed base; this interpretation is also supported by the monotonic decline of the flux (Sect. \ref{sec:vari})

\subsection{Flux decline and the young ejecta scenario}\label{sec:vari}

Fig. \ref{fig:2} displays the radio spectra of J1017 from 0.14 to 4.926 GHz. Over the last decade, it has been firmly detected in several radio surveys, including the LOFAR Two-metre Sky Survey DR2 (LoTSS) at 0.14 GHz \citep[][]{2022A&A...659A...1S}, the Rapid ASKAP Continuum Survey (RACS) at 0.89, 1.37, and 1.66 GHz \citep[][]{2020PASA...37...48M,2021PASA...38...58H,2023PASA...40...34D,2024PASA...41....3D,2025PASA...42...38D}, and the Very Large Array Sky Survey (VLASS) at 3 GHz (\citealt{2020PASP..132c5001L}).

The multi-epoch and multifrequency radio data over $\sim$ 30 years reveal J1017's activity history. This dwarf--dwarf merger was undetected in the NRAO VLA Sky Survey (NVSS) and Radio Sky at Twenty-centimeters (FIRST) surveys at 1.4 GHz in the 1990s \citep{1998AJ....115.1693C,1995ApJ...450..559B}, placing $3\sigma$ upper limits of $\sim 1.5$ mJy and $\sim 0.5$ mJy, respectively (see Table \ref{tab:survey-flux}). However, the recent RACS-mid detection of 3.46 mJy beam$^{-1}$ at 1.37 GHz is well above both upper limits. This indicates that J1017 has brightened significantly over the past $\sim$ 30 years and suggests that the radio emission has been recently triggered. The first radio detection occurred in the LoTSS survey, which implies that episodic accretion triggered jet activity $\gtrsim 10$ years ago; this is consistent with a young transient rather than a steady radio-emitting AGN. We note that J1017 was previously classified as a variable source by \cite{2025ApJ...978..158E} based on a subset of these archival data. J1017 has faded since 2019 or earlier. Using three-epoch VLASS 3 GHz Quick Look data, we detected a declining flux density dropping from 1.52 mJy to 0.84 mJy within five years. We note that VLASS Quick Look images generally have large uncertainties. To address this, we included the systematic errors in our variability analysis. We performed a weighted $\chi^{2}$ test \citep[e.g.,][]{2014MNRAS.438..352B}, which shows a variability with $\chi^{2}\approx7.8$ and a $p$-value $\approx0.02$.

To quantify the fading of J1017, we performed a fit to the multi-epoch and multifrequency radio flux with a temporal spectrum model ($S(\nu,t)\propto\nu^{\alpha}e^{-(\Delta t/\tau)}$; see details in Appendix \ref{app:model}). The best-fit parameters are a spectral index $\alpha=-1.15\pm0.15$ and a decay timescale $\tau=3066\pm1952$ days ($\sim8$ years). This exponential timescale quantifies the fading of the transient emission and predicts that the flux density of J1017 at 5 GHz will drop to $<0.1$ mJy in $\gtrsim12$ years. The dashed lines in Fig. \ref{fig:2} show this best-fit temporal evolution. The declining flux density implies that the young ejecta is expanding and its electrons are cooling. This young ejecta scenario also naturally explains the low flux density at 0.14 GHz in 2015 (LoTSS; pink diamond in Fig. \ref{fig:2}). If the ejecta had been optically thin during the LoTSS observation, our derived steep spectrum would predict an extrapolated 0.14 GHz flux orders of magnitude higher than the observed $\sim 1.2$ mJy. This suggests that the newly born ejecta was more compact, so its low-frequency emission was likely absorbed by synchrotron self-absorption \citep[e.g.,][]{2021MNRAS.503.3886Y,2023A&A...670A..22F}. However, we cannot rule out the possibility that the 2015 detection captured the rising phase of the episodic ejection before its peak. In either case, the observed radio evolution points to a unified physical scenario: a recently triggered ejecta that expands with time, transitioning from an initial compact, self-absorbed phase into a cooling, optically thin plasma.

\subsection{Accretion states and episodic jets in IMBHs}

The fundamental plane (FP) relation, linking black hole mass and radio and X-ray luminosities, is well established for stellar-mass black holes and SMBHs \citep{2003MNRAS.345.1057M,2004A&A...414..895F} but remains poorly constrained for IMBHs due to sparse, heterogeneous samples \citep[e.g.,][]{2022MNRAS.516.6123G,2025ApJ...980...97Y}. According to the standard FP from \cite{2003MNRAS.345.1057M}, with its X-ray luminosity of $\lx=10^{42.2} \ergs$, we would expect a radio luminosity of $\lr=10^{36.8} \ergs$ for J1017. This estimate is more than one order of magnitude lower than our EVN result, indicating that the target does not follow the standard FP. The time interval between our EVN observation and the X-ray observation, and the decaying trend in the radio activity, also contribute to this difference. This deviation indirectly partly supports the interpretation that the system features optically thin ejecta at a high Eddington ratio of $\lx/\ledd\sim0.1$, a state that falls outside the standard FP.

Among the six currently known VLBI-detected IMBHs or candidates \citep{Wrobel2006ApJ...646L..95W,2022MNRAS.514.6215Y,2022ApJ...941...43Y,2023MNRAS.520.5964Y}, J1017 and SDSS J090613.77+561015.2 \citep{2023MNRAS.520.5964Y} are the two most radio-luminous sources (see Fig. \ref{fig:FP} for the $\lr/\ledd-\lr/\ledd$ distribution). Both sources have a radio luminosity ($\lr$) of $\sim 10^{38} \ergs$ ($\lr/\ledd>10^{-6}$). \cite{2023MNRAS.520.5964Y} found that SDSS J090613.77+561015.2 exhibits episodic, relatively large-scale two-sided jet activity. J1017 instead shows a compact, milliarcsecond scale morphology, suggesting an earlier stage of similar jet activity. Notably, J1017 has a distinct accretion state: at $\lx/\ledd\sim0.1$, it is the most luminous of all the VLBI-detected IMBHs. Such a high X-ray luminosity could be a signature of the merger, where efficient gravitational inflow drives gas inflows toward the nucleus \citep[e.g.,][]{2007MNRAS.380..877S,2025ARA&A..63..379K}. In X-ray binaries, high accretion rates can be associated with the transition state, which is often accompanied by discrete ejections \citep{Fender2004MNRAS.355.1105F}. Assuming scale-invariant accretion physics, this physical picture could represent a possible analog of the young ejecta scenario discussed in Sect. \ref{sec:vari}. We further place J1017 in the broader context of local dwarf galaxies. Though luminous among VLBI-detected sources, its radio luminosity is typical of broader dwarf galaxy samples \citep[e.g.,][]{2020ApJ...888...36R,2022MNRAS.516.6123G,2024ApJ...974...66P,2025ApJ...978..158E}, indicating it is not an extreme outlier.

\section{Conclusions}\label{sec:5}

To probe the nature of the radio emission in J1017, a nearby dwarf--dwarf merger system hosting an AGN driven by an IMBH candidate ($M_{\rm BH} \sim 10^{5.3}M_\odot$), we carried out deep EVN observations at 4.926 GHz. We detected a compact radio component on milliarcsecond  scales with a luminosity of $1.39\times10^{38} \ergs$ and a brightness temperature of $5.9\times 10^6$ K at 4.926 GHz, which confirm its nonthermal origin. The radio component is near the \textit{Gaia} centroid and has undergone brightening and fading over the past 30 years. Based on this and its steep spectrum and declining flux trend, we interpret this milliarcsecond-scale radio component as a young, short-lived ejecta.

\bibliography{bibtex/ref}
\begin{appendix}
\nolinenumbers
\FloatBarrier

\section{Acknowledgements}
\begin{acknowledgements}
      The authors thank the anonymous referee for their helpful comments and questions during the review process. This work is supported by the Tianshan Talent Training Program (grant No. 2023TSYCCX0099). N.C. and L.C.H. acknowledge support from the Xinjiang Tianchi Talent Program. L.C.H. is supported by the National Science Foundation of China (12233001) and the China Manned Space Program (CMS-CSST-2025-A09). This work is also partly supported by the National Key R\&D Program of China (grant No. 2024YFA1611500), the Urumqi Nanshan Astronomy and Deep Space Exploration Observation and Research Station of Xinjiang (XJYWZ2303), and the Central Guidance for Local Science and Technology Development Fund (grant No. ZYYD2026JD01).

      The European VLBI Network is a joint facility of independent European, African, Asian, and North American radio astronomy institutes. Scientific results from data presented in this publication are derived from the following EVN project code: EY051. This work has made use of data from the European Space Agency (ESA) mission \textit{Gaia} (\url{https://www.cosmos.esa.int/gaia}), processed by the \textit{Gaia} Data Processing and Analysis Consortium (DPAC; \url{https://www.cosmos.esa.int/web/gaia/dpac/consortium}). Funding for the DPAC has been provided by national institutions, in particular the institutions participating in the \textit{Gaia} Multilateral Agreement.

      LOFAR data products were provided by the LOFAR Surveys Key Science project (LSKSP; \url{https://lofar-surveys.org/}) and were derived from observations with the International LOFAR Telescope (ILT). LOFAR \citep{2013A&A...556A...2V} is the Low Frequency Array designed and constructed by ASTRON. It has observing, data processing, and data storage facilities in several countries, which are owned by various parties (each with their own funding sources), and which are collectively operated by the ILT foundation under a joint scientific policy. The efforts of the LSKSP have benefited from funding from the European Research Council, NOVA, NWO, CNRS-INSU, the SURF Co-operative, the UK Science and Technology Funding Council and the J\"ulich Supercomputing Centre.

      This scientific work uses data obtained from Inyarrimanha Ilgari Bundara, the CSIRO Murchison Radio-astronomy Observatory. We acknowledge the Wajarri Yamaji People as the Traditional Owners and native title holders of the Observatory site. CSIRO’s ASKAP radio telescope is part of the Australia Telescope National Facility (\url{https://ror.org/05qajvd42}). Operation of ASKAP is funded by the Australian Government with support from the National Collaborative Research Infrastructure Strategy. ASKAP uses the resources of the Pawsey Supercomputing Research Centre. Establishment of ASKAP, Inyarrimanha Ilgari Bundara, the CSIRO Murchison Radio-astronomy Observatory and the Pawsey Supercomputing Research Centre are initiatives of the Australian Government, with support from the Government of Western Australia and the Science and Industry Endowment Fund. 

      The National Radio Astronomy Observatory and Green Bank Observatory are facilities of the U.S. National Science Foundation operated under cooperative agreement by Associated Universities, Inc.
\end{acknowledgements} 

\FloatBarrier

\section{Radio flux densities of J1017 from multiple surveys}\label{app:flux_extraction}

We list the radio flux densities of J1017 from multiple surveys in Table \ref{tab:survey-flux}. For the flux extraction, we forced the 2D Gaussian sizes to match their synthesized beams, since the target remains unresolved in these survey images.

\begin{table}[ht!]
\centering
\caption{Radio flux densities of J1017 from multiple surveys.}
\label{tab:survey-flux}
\begin{tabular}{cccc}
\hline
\hline 
Survey   & Date   & $\nu_{\rm obs}$    & Flux    \\
Name     &        & (GHz)   & (mJy beam$^{-1}$)\\
\hline
LoTSS    & 2015-07-08& 0.14 & 1.19$\pm$0.15    \\
RACS-low & 2019-04-21& 0.89 & 4.00$\pm$0.47    \\
RACS-mid & 2021-01-01& 1.37 & 3.46$\pm$0.39    \\
RACS-high& 2021-12-24& 1.66 & 1.83$\pm$0.24    \\
VLASS-ep1& 2019-04-26& 3    & 1.52$\pm$0.21    \\
VLASS-ep2& 2021-12-06& 3    & 1.35$\pm$0.19    \\
VLASS-ep3& 2024-08-03& 3    & 0.84$\pm$0.16    \\
NVSS     & 1993-12-15& 1.4  & $<1.46^{*}$       \\
FIRST    & 1994-08-19& 1.4  & $<0.47^{*}$       \\
\hline
\end{tabular}
\tablefoot{Asterisks indicate $3\sigma$ upper limits for non-detections. A systematic error of 10 percent is included in the error budget of flux density.}
\end{table}
\FloatBarrier

\section{Temporal spectrum model $S(\nu,t)$}\label{app:model}

Given the unknown trigger time and the fading start time of the radio activity, we adopted an exponential decay rather than the power-law models \citep{van1966Natur.211.1131V}, which require a specific epoch. The decay model is as follows:\begin{equation}
    S(\nu,t)\propto\nu^{\alpha}e^{-(\Delta t/\tau)},
\end{equation}
where $\tau$ is the decay timescale, $\Delta t=t-t_{\rm ref}$. We adopted a time-independent spectral index ($\alpha$) in our model since the $\gtrsim1$ GHz emission indicates an optically thin regime, which is expected to remain temporally stable within a narrow frequency range ($\sim1-5$ GHz). Furthermore, the calculated spectral indices from the two epochs, $-0.79\pm0.15$ (2019-04) and $-0.51\pm0.32$ (2021-12), are consistent within $1\sigma$, also confirming this stability. 

We used RACS (0.89 GHz data were excluded due to the possible peak frequency), VLASS, and EVN data as the fitting data. We used the Markov chain Monte Carlo method with \texttt{emcee} \citep{Foreman2013PASP..125..306F}. During the fitting process, we normalized the parameters to a reference frequency of $\nu_{\rm ref}=3$ GHz and a reference epoch of $t_{\rm ref} = 58599$ (the first VLASS epoch) to minimize parameter covariance.

\section{The $\lr/\ledd-\lx/\ledd$ plane}

\begin{figure}[!ht]
    \centering
    \includegraphics[width=0.45\textwidth]{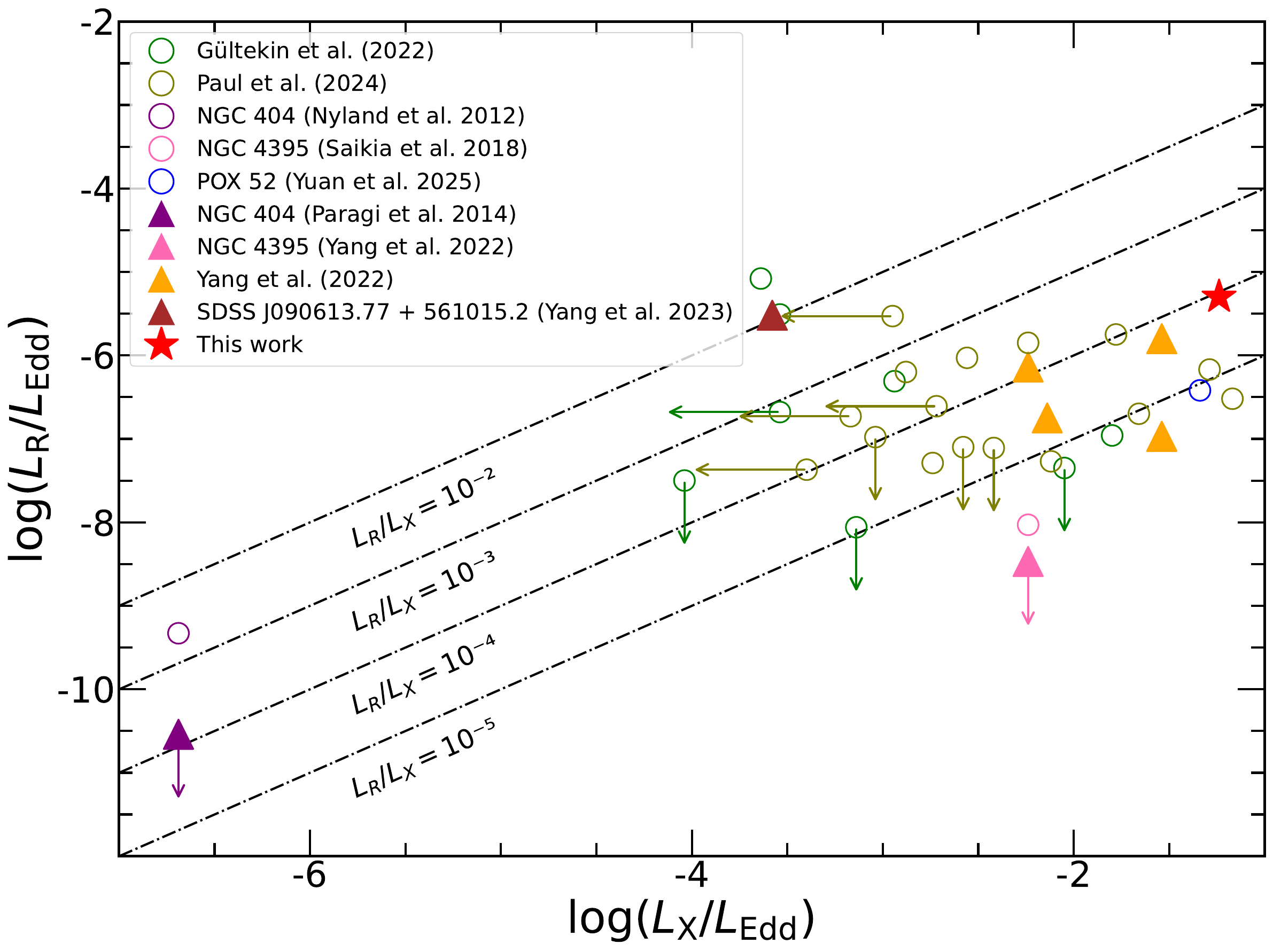}
    \caption{ $\lr/\ledd-\lx/\ledd$ plane. The Eddington luminosity ($\ledd$), 4.926 GHz radio luminosity ($\lr$), and 2$-$10 keV X-ray luminosity ($\lx$) are shown on a logarithmic scale. The dot-dashed black lines indicate different $\lr/\lx$. References in the legend indicate the sources of radio luminosity data. Triangles denote radio luminosity measurements obtained using VLBI. The red star marks J1017 (this work). Arrows indicate that the corresponding data points are upper limits.}
    \label{fig:FP}
\end{figure}

To show jet efficiency under specific accretion states of IMBHs, we plot the $\lr/\ledd-\lx/\ledd$ plane in Fig. \ref{fig:FP}. We collected seven VLBI observed IMBHs from the literature to give a comparison with J1017 \citep{Wrobel2006ApJ...646L..95W,2014ApJ...791....2P,2022MNRAS.514.6215Y,2022ApJ...941...43Y,2023MNRAS.520.5964Y}. We also included Very Large Array (VLA) results for some IMBHs or candidates \citep{2012ApJ...753..103N,2018A&A...616A.152S,2022MNRAS.516.6123G,2024ApJ...974...66P,2025ApJ...980...97Y}.

\end{appendix}
\end{document}